# Silver particles with rhombicuboctahedral shape and effectively isotropic interactions with light

Anja Maria Steiner,[1,2,°] Martin Mayer,[1,2,°] Daniel Schletz,[1] Daniel Wolf,[3] Petr Formanek,[1,4] René Hübner,[5] Martin Dulle,[6] Stephan Förster,[6] Tobias A.F. König,[1,2,7,*] and Andreas Fery[1,2,7,*]

°       contributed equally.
[1]     Leibniz-Institut für Polymerforschung Dresden e.V., Institute of Physical Chemistry and Polymer Physics, Hohe Str. 6, 01069 Dresden, Germany
[2]     Cluster of Excellence Center for Advancing Electronics Dresden, Technische Universität Dresden, 01062 Dresden, Germany
[3]     IFW Dresden, Institute for Solid State Research, Helmholtzstr. 20, 01069 Dresden, Germany
[4]     Helmholtz-Zentrum Dresden-Rossendorf, Institute of Resource Ecology,
[5]     Institute of Ion Beam Physics and Materials Research, Helmholtz-Zentrum Dresden-Rossendorf, Bautzner Landstraße 400, 01328 Dresden, Germany
[6]     JCNS-1/ICS-1, Forschungszentrum Jülich GmbH, Wilhelm-Johnen-Straße, 52428 Jülich
[7]     Department of Physical Chemistry of Polymeric Materials, Technische Universität Dresden, Hohe Str. 6, 01069 Dresden, Germany

Corresponding authors: koenig@ipfdd.de, fery@ipfdd.de

## Keywords

# Abstract

Truly spherical silver nanoparticles are of great importance for fundamental studies including plasmonic applications, but the direct synthesis in aqueous media is not feasible. Using the commonly employed copper-based etching processes, isotropic plasmonic response can be achieved by etching well-defined silver nanocubes. Whilst spherical like shape is typically prevailing in such processes, we established that there is a preferential growth towards silver rhombicuboctahedra (AgRCOs), which is the thermodynamically most stable product of this synthesis. The rhombicuboctahedral morphology is further evidenced by comprehensive characterization with small-angle X-ray scattering in combination with TEM tomography and high resolution TEM. We also elucidate the complete reaction mechanism based on UV-Vis kinetic studies, and the postulated mechanism can also be extended to all copper-based etching processes.

Nowadays, the controlled synthesis of metallic nanoparticles plays an important role in various diverse fields such as optics,[1] sensing,[2] photocatalysis,[3] and therapeutics.[4] Most of these applications rely on the plasmonic effects exhibited by the metallic nanoparticles. In recent years, the focus in this area has shifted significantly into the collective effects of plasmonic nanoparticles resulting in emergent properties,[5] *e.g.,* complex coupling of particle assemblies,[6-8] collective interactions[9, 10] and the interplay of plasmonics with classical photonic properties.[11]

Well-defined nanoparticles as the building blocks in such complex systems are essential. The uniformity and control in terms of size, morphology, and dispersity on the individual level are fundamental to reduce the intrinsic losses of plasmonics (*i.e.*, narrow bandwidths).[12] Especially, in regard of particle assemblies, the precise positioning in traps and finally, the interparticle distances are of great importance.[13] Consequently, the particle geometry and its distribution is crucial to reduce defects in the resulting plasmonic crystal.

Although plasmonic field enhancement is most pronounced at the tips and edges of the anisotropic nanoparticles (such as rods and cubes), the use of spherical metal nanoparticles is indispensable to fundamental studies, since their plasmonic response is isotropic. For example, complex coupling behavior between spherical colloids can be deconvolved to comprehensible models, since the complexity of the individual building block is reduced.[6] Hence, the focus of such system lies solely on the effects of plasmonic coupling between isotropic colloids.[14] Colloidal approaches towards optically isotropic gold nanoparticles have been well-established in a vast range of synthesis protocols.[15-18] However, the plasmonic effect of gold colloids is limited to wavelengths above 515 nm, due to its interband gap transition.[19] In contrast, silver nanoparticles have the capacity to extend over the complete visible optical spectrum (>330 nm),[20] and exhibite a remarkable quality factor in the spectrum range that is inaccessible for gold.[5, 21] Since this optical range complements the absorption band gap of typical metal oxides (*e.g.,* $TiO_2$), the plasmonics of silver can also boost their intrinsic photocatalytic activity.[22, 23]

However, the controlled synthesis of isotropic silver nanoparticles in aqueous media is challenging in terms of shape, size, and dispersity. This is due to the accelerated formation of low-index facets during the growth of silver nanoparticles, resulting in anisotropic nanoparticles with sharp corners.[24-26] Therefore, commonly used synthetic pathways such as controlling kinetics, tuning reduction potentials, and capping facets do not result in the targeted isotropic morphology.[27] To overcome this limitation, additional etching processes such as oxidative etching have been reported to form high-index facets after the initial nanoparticle growth.[8, 28, 29] Most of these studies investigated the reshaping of anisotropic silver colloids using etching agents like copper or iron salts.[28, 30-32] Nevertheless, there is a lack of in-depth studies into the etching mechanism and especially, the formation of facets as a result of the etching mechanism.

In this work, we aim to investigate the etching mechanism of silver nanoparticles in a controlled manner and the formation of new shape and facets during the process. Well-defined silver nanocubes (AgNCs) with a narrow size distribution are etched to isotropic plasmonic silver nanorhombicuboctahedra (AgRCOs) with the use of a catalyst. Copper nitrate is used as the catalyst to ensure controlled and mild reaction conditions during the selective etching of edges. The obtained nanorhombicuboctahedra crystal shape is confirmed by a combination of transmission electron microscopy (TEM) tomography[33] and small-angle X-ray scattering (SAXS) that allows a comprehensive three-dimensional (3D) reconstruction of the morphology and crystallinity of the investigated AgRCOs. The kinetic evolution and plasmonic profile from nanocubes to nanorhombicuboctahedra was studied by UV–vis spectroscopy and accompanied by numerical simulations. Most importantly, we also propose a mechanism based on these results, that involves not only the etching process but also a simultaneous overgrowth process, resulting in the observed equilibrium reaction.

In general, the direct synthesis of spherical or typically quasi-spherical silver nanoparticles is quite limited, since the growth of silver crystals is enforcing the formation of low-index facets.[34] The thermodynamically most stable {100} facets are typically dominating the crystal growth of silver and higher-index facets are only possible under specific synthesis conditions, for example for particles smaller than 30 nm in polyol reactions.[35] Thus, in order to achieve silver nanoparticles with high sphericity, *i.e.* particles with an isotropic optical response, a multistage synthesis route has to be employed. As schematically shown in Fig. 1a (left) and supported by TEM images (Fig. 1b), the induced epitaxial overgrowth of single-crystalline gold spheres leads to the formation of silver nanocubes (AgNCs) of same crystallinity. As recently published, well-defined gold nanoparticles with a narrow size distribution can be exploited for seed-mediated silver overgrowth to ensure controlled shape and dispersity under aqueous conditions.[24, 34] In this protocol, the capping of facets with chloride that was provided by the stabilizing surfactant, thermodynamically enforces six {100} facets with uniformly distributed facet sizes (nanocube structure).

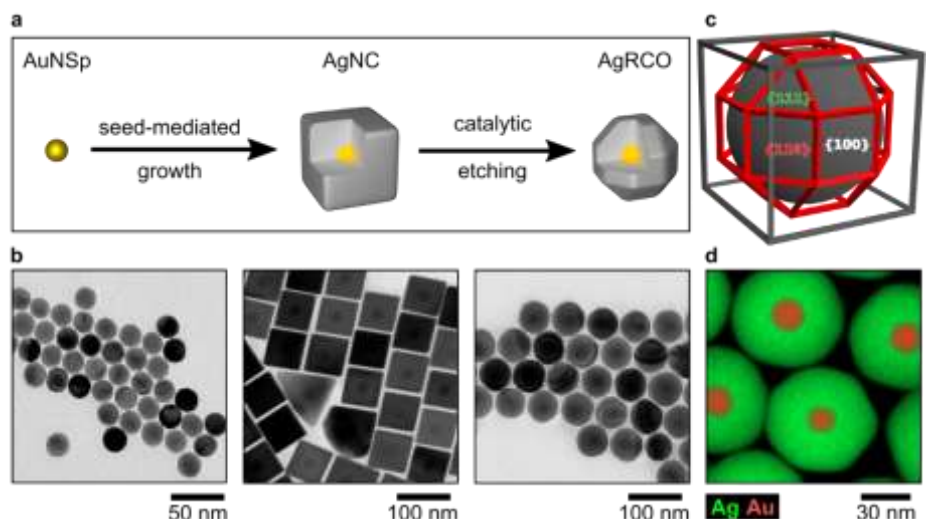

**Figure 1: Synthetic pathway to rhombicuboctahedrically shaped nanoparticles.** (a) First, synthesis of AgNC by facet-selective silver overgrowth of single-crystalline spherical AuNSp. Second, controlled catalytic etching of the AgNCs to yield monodispersed AgRCOs. (b) TEM images of the explicit intermediate particles (left: AuNSp; middle: AgNCs) and the final AgRCOs (right). (c) 3D sketch of a rhombicuboctahedron to visualize the particle geometry. (d) EDX element distribution of silver and gold in the final AgRCOs.

Following the growth of AgNCs, a catalytic etching step (Fig. 1a, right) is applied to these nanocubes to overcome the limitation of low-index facets and to achieve a final spherical morphology. In this case, copper nitrate is employed as the catalyst because it produces the required etching agent in an *in-situ* manner under mild catalytic conditions,[28] which is crucial to control the final morphology and to avoid additional polydispersity. Within this etching process, hydroxide radicals are formed and act as the actual etching agent by oxidizing the silver particles (Ag(0) → Ag(I)). The reaction sub-equations of the etchant formation are summarized in the Supporting Information S1 and the overall mechanism will be discussed later.[36-38] As shown in Fig. 1d, the etched silver nanoparticles still exhibit defined facets, leading to preferable oxidation of corners and edges by the etchant. Typically, etching processes aim to achieve a structure exhibiting minimal surface area, *i.e.* a sphere. However, the intrinsic surface energies of silver prefer the formation of low-index {100} and {110} facets and lead to the formation of a rhombicuboctahedral morphology (AgRCOs), as illustrated in Fig.1c. In this geometry, the overall surface area is minimized through the etching process, while the surface area of the low-index {100} facets is maintained as high as possible (for detailed discussion, see Supporting Information Fig. S1).

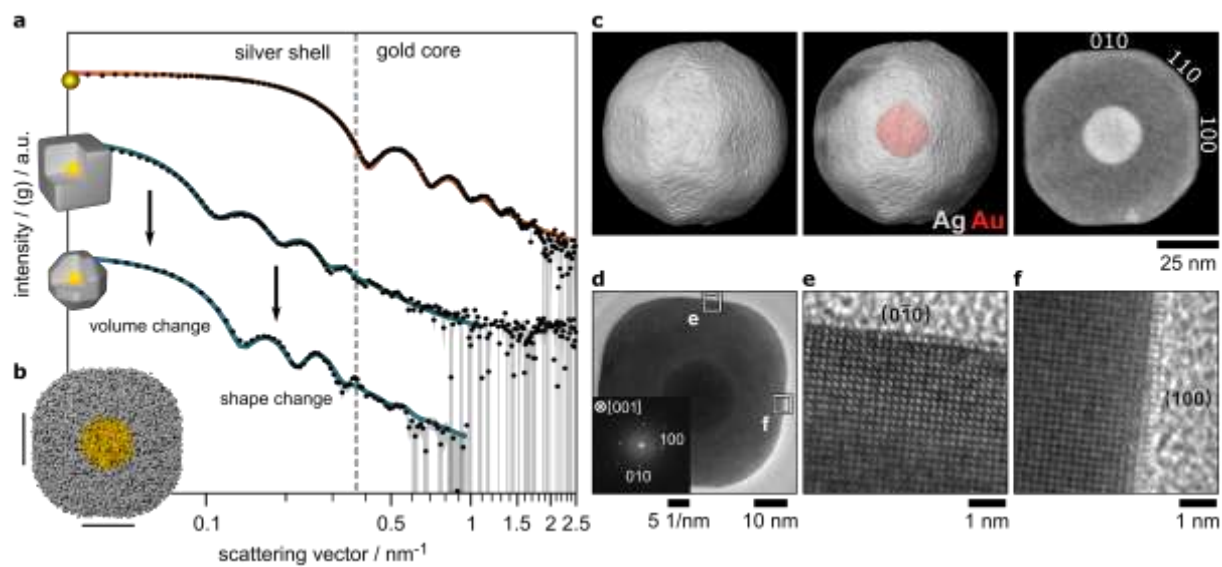

**Figure 2: Morphology of the AgRCOs *via* SAXS and TEM tomography.** (a) SAXS investigation of the intermediate particles and the final AgRCOs (dots). Best-fit modeling of the SAXS data to deconvolute the shape from the ensemble (orange: analytical fit; blue: numerical fit). (b) 3D reconstruction from SAXS model revealing the formation of flattened facets.(c) 3D iso-surface rendering of the reconstructed TEM tomogram to determine the rhombicuboctahedral shape and 2D slice through the center indicating the formation of defined facets (right), as predicted from SAXS fitting. (d-f) HRTEM investigation of the crystal structure of the final rhombicuboctahedron. (e,f) insets of (d).

Since the quasi-spherical shape cannot be completely excluded by TEM imaging (see Fig.1b, right), further characterization of the morphology is required to confirm the AgRCOs shape. Firstly, the uniformity and control in terms of size, morphology, and dispersity of the silver colloids are evaluated statistically by SAXS (see Supporting Information for more details). Fig. 2a depicts the scattering profiles of the different intermediate synthesis steps using SAXS. Starting from the initial spherical gold particles, the analytical model reveals a diameter size of $(21.5 \pm 0.8)$ nm. For the AgNCs and rhombicuboctahedra (AgRCOs), numerical modelling and fitting of SAXS profiles are performed to reconstruct a comprehensive 3D shape from the experimental SAXS data (for more details, see Schnepf *et al.* [39] and Supporting Information S2). The obtained AgNCs have an edge length of $(65.0 \pm 2.3)$ nm with edge rounding below the detection limit (<5%). Following the etching process, the volume change and the reshaping can be tracked by the scattering vector ($q$) in the respective range (see Fig. 2a). The reconstructed 3D model, as shown in Fig. 2b, has a size / diameter of $(60.0 \pm 2.1)$ nm. Additionally, the presence of planar faces (indicated in Fig 2b) in combination with an edge rounding of $35\% \pm 2\%$ suggests a rhombicuboctahedral shape. Furthermore, the polydispersity of all particles stays nearly constant (3.7%, 3.5%, and 3.5% for the AuNSps, AgNCs, and AgRCOs, respectively), as evident from the SAXS data.

TEM tomography on an individual representative particle was performed to have a closer look into the AgRCOs facet formation. The reconstructed tomogram (Fig. 2c) exhibits the discussed rhombicuboctahedral shape with a distinct gold core. As described elsewhere, the gold core plays an important role within the synthesis and defines the crystallinity.[24, 25, 34] Thus, the final particle is enclosed by the defined {100} and {110} facets on the sides and {111} facets at the edges (see also Fig. 1c, Supporting Information Fig. S3 and Supporting Video SV1), leading to a rhombicuboctahedral shape.

In order to proof the rhombicuboctahedron morphology of the AgRCOs observed by TEM tomography and SAXS, we performed high resolution TEM (HRTEM) imaging (see Fig. 2d–f). This enables us to map the atomic structure of the particles simultaneously with their (projected) shape, visible as edges corresponding to their crystallographic facets (for more details, see Supporting Information S2 and Fig. S4).

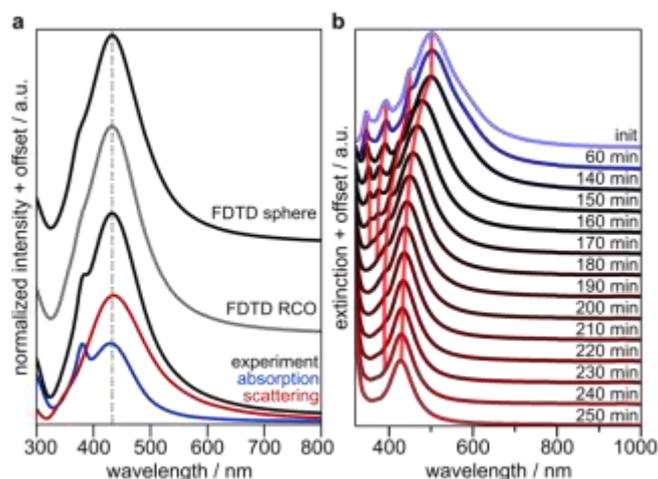

**Figure 3: Plasmonic properties of and evolution to AgRCOs.** (a) Direct comparison of FDTD simulations of perfect spheres and rhombicuboctahedra, and experimental DRA measurements to determine the contribution of scattering (red) and absorption (blue) in the extinction (black). (b) Spectral evolution (guide to the eye for all relevant modes, red) due to the time-dependent selective etching of the AgNCs (init) to AgRCOs (210 min) and slow decaying of the etching rate (LSPR shift/min).

As mentioned above, the morphology of the etched silver nanoparticles is not perfectly spherical. However, the AgRCOs behave optically like spherical nanoparticles of the same size, since plasmonics are dominated by sharp features. As supported by numerical simulations in Fig. 3a, spheres and RCO display almost identical plasmonic profiles. Consequently, the spectral position of the experimentally measured plasmonic modes can be matched with the simulated results for both RCO and sphere. In direct comparison of experiment and simulations, the narrow size distribution of the experimentally realized RCOs results in a plasmonic bandwidth close the theoretical limit. Further deconvolution of the absorption and scattering profile obtained experimentally also reveals a narrow size distribution of the nanoparticles and the high quality factor of silver.

To have an in-depth insight into the mechanism, the described etching process can be followed using UV-Vis spectroscopy. Due to the high sensitivity of plasmonics with corners and edges, minimal changes during the transition from AgNCs to AgRCOs can be tracked from the spectral evolution, as demonstrated in Fig. 3b. For the initial AgNCs, the predominant dipolar mode is located at 505 nm, and higher-order plasmonic modes can be observed in the range from 350 nm to 450 nm, which indicates low dispersity and the presence of sharp corners in AgNCs. [25] As soon as the etching of the AgNCs starts, higher-order modes related to the edges of AgNCs disappear and the dipolar modes also shift to lower wavelengths (Fig. 3b, guide to the eye, red).The loss of higher modes[40] can be attributed to the increment of isotropy,[41] and the observed blue shift is contributed by the etching of corners and edges.[42] After approximately 3.5 h (Fig. 3a, 210 min), uniform AgRCOs are reached and exhibit a single plasmonic mode at 439 nm wavelength. Reaching the spectral range with a high quality factor for silver, the dipolar mode narrows (the full with half maximum (FWHM) from 119 nm at 0 min to 79 nm at 210 min) and the quadrupolar mode emerges at 372 nm. As the reaction continues, the etching of silver will slowly decay until all of the ascorbic acid (AscH$_2$) is consumed. The proposed mechanism and the optical evolution can be also transferred to other particle sizes (see Supporting Information Fig. S5).

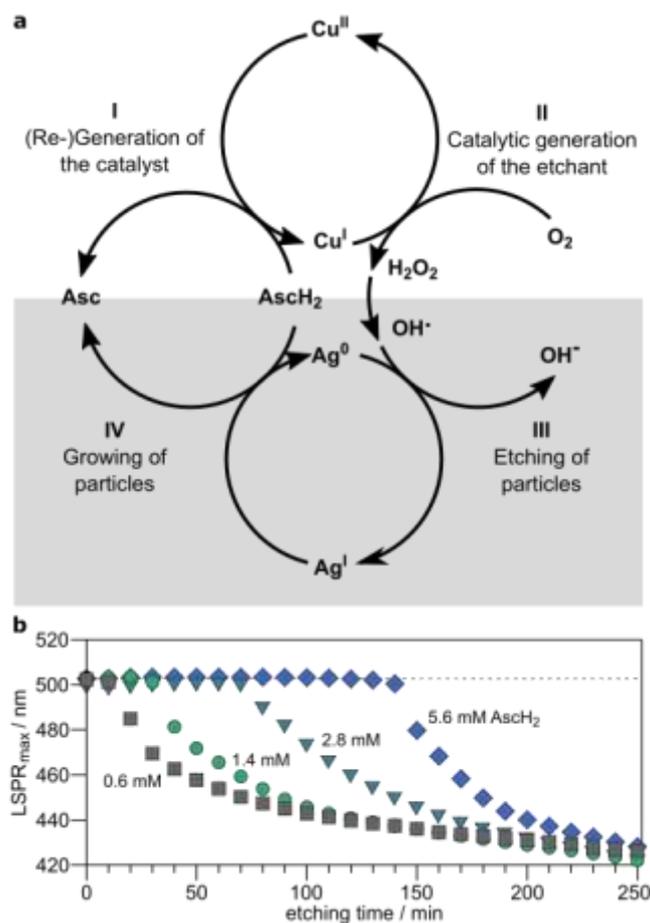

**Figure 4: Simplified representation of the catalytic etching mechanism.** (a) First, the catalyst Cu(I) is generated (step I). This in turn is needed in the second step to form the etchant (OH·, formed from dissolved O$_2$, with H$_2$O$_2$ as intermediate product) which etches the AgNCs (step III). The resulting Ag(I) reacts with the ascorbic acid (AscH$_2$, step IV) which has been added to the solution to (re-)generate the catalyst in the first step. Balancing the reactions of etching and simultaneous growth, the initial cubically shaped particles are selectively etched to RCO. (b) Kinetic evolution of the dipolar LSPR wavelength upon change of the AscH$_2$ concentration. Decrease of AscH$_2$ results in earlier adjustment of the equilibrium concentration and slow down of the actual etching rate (LSPR shift per minute).

To apprehend and formulate the complete mechanism behind the growth of AgRCOs, the interplay of all reactions has to be considered. Fig. 4a provides an overview of the four main reactions occurring during the etching process. In the first step, the active species of the catalyst is generated by the reduction of Cu(II) to Cu(I) by ascorbic acid (AscH$_2$). As confirmed in literature,[36] the actual etchant, consisting of hydroxyl radicals (step II), is formed through a cascade of reactions (see Supporting Information S1). In presence of dissolved oxygen, the active catalyst is oxidized and dioxid(−)-ions are formed. The superoxide in turn reacts to oxygen peroxide. Next, the hydrogen peroxide initiates the etching process by forming hydroxide radicals in presence of the active catalyst. The Ag(0) is then oxidized to Ag(I) ions in Step III under the presence of the hydroxide radicals and dissolved by forming a complex with the stabilizing surfactant (hexadecyltrimethylammonium chloride; CtaC). However, the presence of the reducing agent $AscH_2$ also leads to a simultaneous facet-selective silver growth in Step IV (see also experimental part of AgNCs). Hence, the actual reshaping and facet formation, leading to AgRCOs, is shown to be an equilibrium between etching (step III) and simultaneous particle growth (step IV), as highlighted in grey in Fig. 4a. To check once more the role of the different reactants, several control experiments were performed. Briefly, the exclusion of dissolved oxygen or ascorbic acid leads to no visible reshaping. Consistent with the role as catalyst, changes of the copper concentration only affect the kinetics but not the final particle shape (see Supporting Information S6).

The existing equilibrium reaction is also evi[5]denced by the previous kinetic studies (Fig. 3b and 4b), where the reshaping of AgNCs is delayed by several hours. Within this period of time, the reduction and subsequent deposition of silver (formed Ag(I)) on the nanoparticles (particle growth) is dominating the reaction. Since the conditions for mild and selective silver growth require an excess of $AscH_2$, its consumption shifts the equilibrium gradually from growth towards etching with time. As shown in Fig. 4b, varying the amount of $AscH_2$ in the system influences the etching process by tuning the equilibrium. Increasing the amount of $AscH_2$ delays the reshaping of the particles because the time span dominated by the growth phase is extended. Thus, the adjustment of the equilibrium takes longer. The $AscH_2$ concentration, however, also has an impact on the overall etching rate (LSPR shift per minute), since the accompanied pH also influences its reduction potential.[43] Therefore, not only the initial etching is delayed, the apparent etching rate (LSPR shift per minute) is also affected by the change in the amount of $AscH_2$.

Since the formation of facet dictated by capping and stabilization specific facets, the presence and identity of ions on the surface of the nanoparticles can provide further information on the overall mechanism. In general, the surfactant headgroup, by comparison of CtaC with benzylhexadecyldimethylammonium chloride (BdaC), has an influence on the reaction kinetics, but not at the formation of specific facets.[44] As shown in the element maps in Supporting Information S7, the surface of the nanoparticles is mainly covered by chloride ions and almost no adsorbed copper is found on the surface. Hence, the formation of the facets is primary forced by the adsorption of chloride on facets, analogous to the well-known silver growth. As described in reference 32, this is supported by the formation of Ag-O complexes at high index facets. In addition to the formation of the etchant, copper is able to drag the oxygen from the surface and makes the high index facets available for etching reactions. Due to the energetically favored passivation of low-index facets and a simultaneous overgrowth, even an etching process is unable to form unfaceted and truly spherical silver colloids.

## Conclusion

In summary, we have demonstrated the formation of thermodynamically most stable silver rhombicuboctahedra (AgRCOs) by exploiting an additional etching step to reduce the isotropy of accessible silver colloids (silver nanocubes). This additional step is essential due to the fact that direct synthesis of narrowly distributed spherical silver particles with high shape-yield is not feasible in pure aqueous media. Although these AgRCOs behave optically analogous to spherical nanoparticles, the resulting rhombicuboctahedral shape is further substantiated by SAXS, TEM tomography, and HRTEM. For the first time, we also elucidated the complete reaction mechanism, which includes a simultaneous growth in addition to the etching process and which is intrinsically linked to all copper-based etching processes. This equilibrium then enforces the formation of the most-isotropic nanoparticles, *i.e.,* rhombicuboctahedra from nanoparticles exhibiting low-index facets. Considering only highly isotropic silver particles can extend the accessible optical range below the bandgap of gold (330–500 nm), the introduced monodisperse RCOs are competent to extend the accessible optical range due to their isotropic plasmonics. In contrast to established quasi-spherical silver systems, our protocols achieves extraordinary narrow bandwidths, which are close to the theoretical limit of perfect spheres. Bringing these AgRCOs to applications, the plasmonic response and therefore, the sensitivity, of the final system is increased due to lack of undesired spectral broadening. Furthermore, unity in particle shape is fundamental for colloidal self-assembly processes, since perfect match between template and colloid is required. This opens up new possibilities in applications based on collective plasmonic coupling such as lasing, [45, 46] photovoltaics,[47] sensing,[48] and higher harmonic plasmonics.[11]

## Acknowledgments

The authors acknowledge the Deutsche Forschungsgemeinschaft (DFG) within the Cluster of Excellence "Center for Advancing Electronics Dresden" (cfaed) for financial support. This project was financially supported by the Volkswagen Foundation through a Freigeist Fellowship, grant no. 92 902, to T.A.F.K. The use of HZDR Ion Beam Center TEM facilities and the support by its staff are gratefully acknowledged. In particular, we acknowledge the funding of TEM Talos by the German Federal Ministry of Education and Research (BMBF), grant no. 03SF0451 in the framework of HECMP. D.W. acknowledges funding from the European Research Council *via* the ERC-2016-STG starting grant atom. We sincerely thank Charlene Ng for valuable discussion of the manuscript.

## Conflict of interest

The authors declare no conflict of interest.